\documentclass[12pt]{article}

\usepackage{scicite}
\usepackage{graphicx} 

\newcommand{\MSUNYR}{M_\odot/yr}

\usepackage{gensymb} 
\usepackage{amssymb}

\usepackage{times}

\newenvironment{sciabstract}{%
\begin{quote} \bf}
{\end{quote}}

\newcounter{lastnote}

\title{Imaging of the CO snow line in a solar nebula analog}

\author
{Chunhua Qi$^{1\ast}$, Karin I. \"Oberg$^{2\ast}$, David J. Wilner$^{1}$,
  Paola d'Alessio$^{3}$, \\Edwin Bergin$^{4}$, Sean
  M. Andrews$^{1}$, Geoffrey A. Blake$^{5}$, \\Michiel R. Hogerheijde$^{6}$, Ewine F. van Dishoeck$^{6,7}$\\
\\
\normalsize{$^{1}$Harvard-Smithsonian Center for Astrophysics}\\
\normalsize{$^{2}$Departments of Chemistry and Astronomy, University of Virginia}\\
\normalsize{$^{3}$Centro de Radioastronomõa y Astrofisica, Universidad Nacional Autonoma de Mexico}\\
\normalsize{$^{4}$Department of Astronomy, University of Michigan}\\
\normalsize{$^{5}$Division of Geological and Planetary Sciences,
  California Institute of Technology}\\
\normalsize{$^{6}$Leiden Observatory, Leiden University}\\
\normalsize{$^{7}$Max Planck Institute for Extraterrestrial Physics}\\
\\
\normalsize{$^\ast$Contributed equally to this manuscript}
}

\date{}

\begin{document}

\maketitle

\begin{sciabstract}
Planets form in the disks around young stars. Their formation
efficiency and composition are intimately linked to the protoplanetary
disk locations of ``snow lines'' of abundant
volatiles. We present chemical imaging of
the CO snow line in the disk around TW~Hya, an analog of the
solar nebula, using high spatial and 
spectral resolution Atacama~Large~Millimeter/Submillimeter~Array
(ALMA) observations of N$_2$H$^+$, a reactive ion present in large
abundance only where CO is frozen out. The N$_2$H$^+$ emission is
distributed in a large ring, with an inner radius that matches CO snow
line model predictions. The extracted CO snow line radius of $\sim30$~AU
helps to assess models of the formation dynamics of the
Solar System, when combined with measurements of the bulk composition
of planets and comets.  

\end{sciabstract}

Condensation fronts in protoplanetary disks, where abundant
volatiles deplete
out of the gas phase and are incorporated into solids, are believed to have 
played a critical role in the formation of planets in the Solar System 
\cite{Lewis74,Stevenson88},
and similar ``snow lines" in the disks around young stars should
affect the
ongoing formation of exoplanets. Snow lines can enhance particle growth and thus 
planet formation efficiencies because of 1) substantial increases in solid mass surface densities 
exterior to snow line locations, 2) continuous freeze-out of gas diffusing across the snow line (cold-head effect), 3) pile-up of dust just inside of the snow line in pressure traps, and 4) an increased stickiness of icy grains 
compared to bare ones, which favors dust coagulation \cite{Ciesla06,Johansen07,Chiang10,Gundlach11,Ros13}. Experiments and theory on these processes have been focused on the H$_2$O snow line, but the results should be generally applicable to snow lines of abundant volatiles, with the exception that the ``stickiness" of different icy grain mantles varies. The locations of snow lines of the most abundant volatiles -- H$_2$O, CO$_2$, and CO  -- with respect to the planet-forming zone may also regulate the bulk composition of planets \cite{Oberg11e}. Determining snow line locations is therefore key to probing grain growth, and thus planetesimal and planet formation efficiencies, and elemental and molecular compositions of planetesimals and planets forming in protoplanetary disks, including the solar nebula.

Based on the Solar System composition and disk theory, the H$_2$O snow line developed at $\sim$3~Astronomical Units (AU, where 1 AU is the distance from the Earth to the Sun) from the early Sun during the epoch
of chondrite assembly \cite{Hayashi81}. 
In other protoplanetary disks the snow line locations are determined
by the disk midplane temperature structures, set by a time dependent combination of the luminosity of the
central star, the presence of other heating sources, the efficiencies of dust and gas cooling, and the intrinsic condensation temperatures of different volatiles. Because of the low condensation temperature of CO, the CO 
snow line occurs at radii of 10's of AU around Solar-type stars: this larger size scale makes 
the CO transition zone the most accessible to direct observations. The
CO snow line is also important in its own right, because CO ice is a
starting point for a complex, prebiotic chemistry
\cite{Herbst09}. Also without incorporating an enhanced grain growth
efficiency beyond that expected for bare silicate dust, observations
of centimeter sized dust grains in disks, including in TW Hya
\cite{Wilner05}, are difficult to reproduce in the outer disk. Condensation of CO is very efficient below the CO
freeze-out temperature, with a sticking efficiency close to unity
based on experiments \cite{Bisschop06}, and a CO condensation-based
dust growth mode may thus be key to explaining these observations. 

Protoplanetary disks have evolving radial and vertical temperature
gradients, with a warmer surface where CO remains in the
gaseous state throughout the disk, even as it is frozen in the cold, dense
region beyond the midplane snow line \cite{Aikawa99}. This means
that the midplane snow line important for planet formation constitutes
a smaller portion of a larger condensation surface.
Because the bulk of the CO emission comes from the disk surface
layers, this presents a challenge for locating the CO midplane snow
line. Its location has been observationally identified
toward only one system, the disk around HD~163296, based on
(sub-)millimeter interferometric observations of multiple CO  
rotational transitions and isotopologues at high spatial resolution, 
interpreted through detailed modeling of the disk dust and gas physical 
structure \cite{Qi11}. An alternative approach to constrain the CO snow line, 
 suggested in \cite{Qi13} and pursued here, is to image molecular emission from a species that 
is abundant only where CO is highly depleted from the gas phase.

N$_2$H$^+$ emission is expected to be a robust tracer of CO depletion
because the presence of gas phase CO both slows down N$_2$H$^+$ formation 
and speeds up N$_2$H$^+$ destruction. N$_2$H$^+$ forms through reactions 
between N$_2$ and H$_3^+$, but most H$_3^+$ will instead transfer a 
proton to CO as long as the more abundant CO remains in the gas phase.
The most important destruction mechanism for N$_2$H$^+$ is proton transfer 
to a CO molecule, whereas in the absence of CO, N$_2$H$^+$ is destroyed 
through a much slower dissociative recombination reaction. These simple
astrochemical considerations predict a correlation between N$_2$H$^+$ and CO 
depletion, or equivalently an anti-correlation between N$_2$H$^+$ and gas-phase CO. 
The latter has been observed in many pre-stellar and protostellar environments, 
confirming the basic theory \cite{Bergin02,Jorgensen04}. In disks N$_2$H$^+$ should therefore be present at large abundances only inside the vertical and horizontal thermal layers where CO vapor is condensing, i.e., beyond the CO snow line.  Molecular line surveys of disks have shown that N$_2$H$^+$ is only present in disks cold enough to entertain CO freeze-out \cite{Oberg11a}, and marginally resolved observations hint at a N$_2$H$^+$ emission offset from the stellar position
\cite{Qi13}, in agreement with the models of disk chemistry
\cite{Walsh12}. Detailed imaging of N$_2$H$^+$ emission 
in protoplanetary disks at the scales needed to directly reveal CO snow lines with
sufficient sensitivity has previously been out of reach.

We used ALMA to obtain images of emission from the 372 GHz dust continuum and
the N$_2$H$^+$ $J=4-3$ line from the disk around TW Hya (Fig.~1, S1)\cite{SOM}. TW Hya is the closest (54$\pm$6 pc) and as such the most intensively-studied pre-main-sequence star with a gas-rich 
circumstellar disk \cite{Kastner97,Qi08}. 
Based on previous observations of dust and CO emission, and the 
recent detection of HD line emission \cite{Bergin13}, this 3--10 million year old, 
0.8 M$_{\odot}$ T Tauri star (spectral type K7) is known to be
surrounded by an almost face-on ($\sim$6$\degree$  
inclination) massive $\sim$0.04~M$_{\odot}$ gas-rich disk. The
disk size in millimeter dust is $\sim$60~AU, with a more extended ($>$
100 AU) disk in gas and micrometer-sized dust \cite{Andrews12}. Both the disk mass and size conforms well with solar nebula estimates -- the minimum mass of the solar nebula is 0.01~M$_{\odot}$ based on planet masses and compositions \cite{Hayashi81} -- and the disk around TW Hya may thus serve as a template for planet formation in the solar nebula. Our images show that N$_2$H$^+$ emission is 
distributed in a ring with an inner diameter
of 0.8 to 1.2 arcsec (based on visual inspection), corresponding to a
physical inner radius of 21 to 32~AU. 
By contrast, CO emission is detected down to radial 
scales $\sim$2~AU \cite{Rosenfeld12a}. The clear difference in 
morphology between the N$_2$H$^+$ and CO emission can be simply explained 
by the presence of a CO midplane snow line at the observed inner edge
of the N$_2$H$^+$ emission ring. The different morphologies cannot be
explained by a lack of ions in the inner disk based on previous spatially
and spectrally resolved observations of another ion, HCO$^+$
\cite{Qi08}. These HCO$^+$ observations had lower sensitivity and
angular resolution 
than the N$_2$H$^+$ observations, but they are sufficient to exclude a 
central hole comparable in size to that seen in N$_2$H$^+$.

To associate the inner edge radius of the N$_2$H$^+$ emission with a midplane temperature, and thus a CO freeze-out temperature,  
requires a model of the disk density and temperature structure.
We adopted the model presented in \cite{Qi08}, updated to conform with recent observations of the accretion rate and 
grain settling (Fig.~S2--S4, Table S1) \cite{SOM}. In the context of this 
disk structure model, the N$_2$H$^+$ inner edge location implies that 
N$_2$H$^+$ becomes abundant where the midplane temperature drops to 16--20~K.
This is in agreement with expectations for the CO freeze-out 
temperature based on the outcome of the laboratory experiments and desorption modeling 
by \cite{Bisschop06}, who found CO condensation/sublimation
temperatures of 16--18~K under interstellar conditions, assuming
heat-up rates of 1~K per 10$^2$ to 10$^6$ years. In outer disk
midplanes, condensation temperatures are expected to
 at most a few degrees higher because of a weak dependence on density
\cite{Hollenbach09}. If CO condenses onto H$_2$O ice rather
than existing CO ice, the condensation temperature will increase further, but this
will only affect the first few monolayers of ice and is not expected
to change the location where the majority of CO freezes out. Some CO may also remain in the gas phase below
the CO freeze-out temperature in the presence of efficient non-thermal
desorption, especially UV photodesorption \cite{Willacy07}, but this
is expected to be negligible in the disk midplane at 30~AU, because of UV shielding by upper disk layers. 
UV photodesorption may
affect the vertical CO snow surface location, however, and it may thus
not be possible to describe the radial and vertical condensation  
surfaces by a single freeze-out temperature.

To locate the inner edge of the N$_2$H$^+$ ring more quantitatively,
we simulated the N$_2$H$^+$ emission with a power-law column
density distribution and compared with the data. We assumed the disk
material orbits the central star in Keplerian motion, and fixed the 
geometric and kinematic parameters of the disk that affect its observed
spatio-kinematic behavior \cite{Qi08}. We used the same, updated density and 
temperature disk structure model \cite{SOM}, and assumed that the  
N$_2$H$^+$ column density structure could be approximated as a radial power-law 
with inner and outer edges, while vertically the abundance was
taken to be constant between the lower (toward midplane) and upper
(toward surface) boundaries.  This approach crudely mimics
the results of detailed astrochemical modeling of disks, 
which shows that molecules are predominantly present in well-defined 
vertical layers \cite{Aikawa99,Walsh12}, and has been used
to constrain molecular abundance structures in a number of 
previous studies \cite{Qi08,Qi13}. 
The inner and outer radii, power-law index, and 
column density at 100~AU were treated as free parameters. We
calculated a grid  of synthetic N$_2$H$^+$ visibility datasets using
the RATRAN code \cite{Hogerheijde00} to determine the radiative transfer and
molecular excitation, and compared with the N$_2$H$^+$ observations.
We obtained the best-fit model by minimizing $\chi^2$, the weighted
difference between the data and the model with
the real and imaginary part of the complex visibility
measured in the ($u,v$)-plane sampled by the ALMA observations of
N$_2$H$^+$. 

Fig.~2a demonstrates that the inner radius 
is well constrained to 
28--31~AU (3$\sigma$). This edge determination was aided by the nearly
face-on viewing geometry, because this minimizes the impact of the
detailed vertical structure on the disk modeling outcome. Furthermore, the Keplerian kinematics of the gas
help to constrain the size scale at a level finer than the spatial
resolution implied by the synthesized beam size. As a result, the fitted inner
radius is robust to the details of the density and temperature
model \cite{SOM} (Table S2). In the context of this model, the best-fit 
N$_2$H$^+$ inner radius corresponds to a CO midplane snow line at a
temperature of 17~K. Fig.~2b presents the best-fit N$_2$H$^+$ column density 
profile together with the best-fit $^{13}$CO profile, assuming a CO freeze-out 
temperature of 17~K \cite{SOM} (Fig.~S5, Table S3). We fit $^{13}$CO
emission (obtained with the Submillimeter Array \cite{SOM}) because the main isotopologue CO lines are optically thick.
The N$_2$H$^+$
column density contrast across the CO snow line is at least an
order of magnitude \cite{SOM}. 
Fig.~2c shows simulated ALMA observations of the best-fit N$_2$H$^+$ $J=4-3$ 
model, demonstrating the excellent agreement.  

Our quantitative analysis thus confirms the predictions that N$_2$H$^+$ traces the snow line of the abundant volatile, CO. Furthermore, the agreement between the quantitative analysis and the visual estimate of the N$_2$H$^+$ inner radius demonstrate 
that N$_2$H$^+$ imaging is a powerful tool to determine the CO snow line 
radii in disks, whose density and temperature structures have 
not been modeled in detail. N$_2$H$^+$ imaging with ALMA may therefore be used 
to provide statistics on how snow line locations depend on parameters of 
interest for planet formation theory, such as the evolutionary stage
of the disks.  

The locations of snow lines in solar nebula analogs like TW Hya are also 
important to understand the formation dynamics of the Solar System. The 
H$_2$O snow line is key to the formation of Jupiter and Saturn
\cite{Lecar06}, while  
CH$_4$ and CO freeze-out enhanced the solid surface density further out in 
the solar nebula, which may have contributed to the feeding zones of Uranus 
and Neptune \cite{DodsonRobinson09}, depending on exactly where these
ice giants formed. In the  
popular Nice model for the dynamics of the young Solar System, Uranus 
formed at the largest radius of all planets, at $\sim17$ AU
\cite{Tsiganis05}, and most  
comets and Kuiper Belt objects formed further out, to $\sim35$ AU. 
The plausibility of this scenario can be assessed using the bulk 
compositions of these bodies together with knowledge of the CO snow line 
location. In particular, Kuiper Belt objects contain CO and the even more 
volatile N$_2$ \cite{Owen93,Tegler12}, which implies that they must have
formed beyond the  
CO snow line. Comets exhibit a range of CO abundances, some of which seem 
to be primordial, which suggest the CO snow line was located in the outer 
part of their formation region of 15--35 AU \cite{Mumma11}. This is
consistent with  
the CO snow line radius that we have determined in the TW Hya disk. However, 
in the context of the Nice model, this CO snow line radius is too large for 
the ice giants, and suggests that their observed carbon enrichment has a 
different origin than the accretion of CO ice
\cite{DodsonRobinson09}. A caveat is that H$_2$O  
ice can trap CO, though this process is unlikely to be efficient enough to 
explain the observations. In either case, the CO snow line locations in 
solar nebula analogs like TW Hya offers independent constraints on the early 
history of the Solar System. 

\begin{figure}%
\begin{center}       
\includegraphics[width=1.0\textwidth]{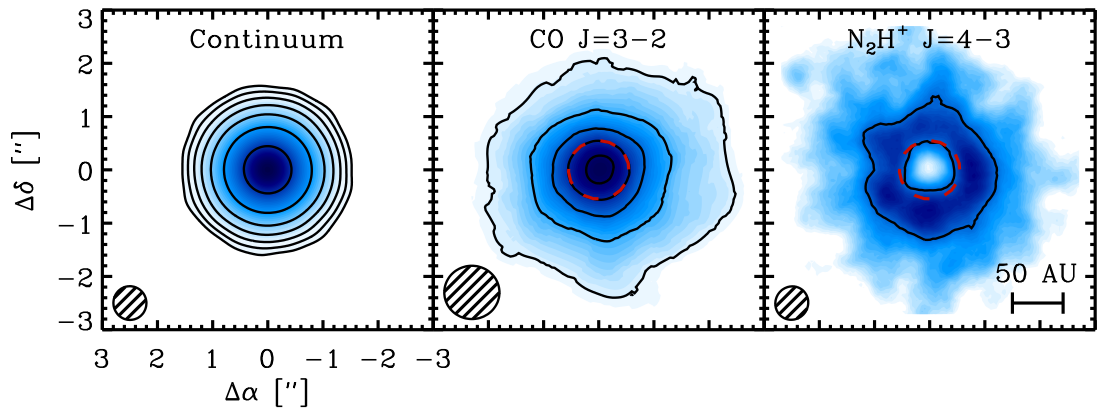}
\caption{Observed images of dust, CO and N$_2$H$^+$ emission toward TW Hya. {\sl Left:} ALMA 372 GHz continuum map, extracted from the 
line free channels of the N$_2$H$^+$ observations. Contours mark [5, 10, 20, 40, 80, 160, 320] mJy beam$^{-1}$ and the rms is 0.2 mJy beam$^{-1}$.
{\sl Center:} image of CO $J=3-2$ emission acquired with the SMA \cite{Andrews12}. Contours mark [1, 2, 3, 4, 5] Jy km s$^{-1}$ beam$^{-1}$ and the rms is 0.1 Jy km s$^{-1}$ beam$^{-1}$. {\sl Right:} ALMA
image of N$_2$H$^+$ $J=4-3$ integrated emission with a single contour at 150 mJy km s$^{-1}$ beam$^{-1}$ and the rms is 10 mJy km s$^{-1}$ beam$^{-1}$. The synthesized beam sizes are shown in the bottom left corner of each panel. 
The red dashed circle marks the best-fit inner radius of the
N$_2$H$^+$ ring from a modeling of the visibilities. This inner edge traces the onset of CO freeze-out according to astrochemical 
theory, and thus marks the CO snow line in the disk midplane.}
\label{mommap}
\end{center}
\end{figure}

\begin{figure}%
\begin{center}       
\begin{minipage}[c]{0.42\linewidth}
	\includegraphics[width=1.0\textwidth]{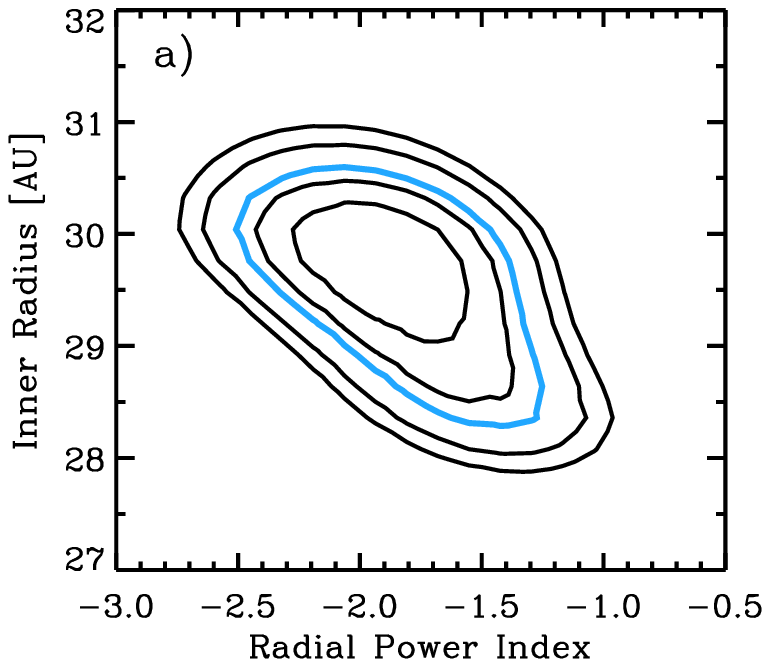}
\vspace{-0.2in}
\end{minipage}\hfill
\begin{minipage}[c]{0.57\linewidth}
	\includegraphics[width=1.0\textwidth]{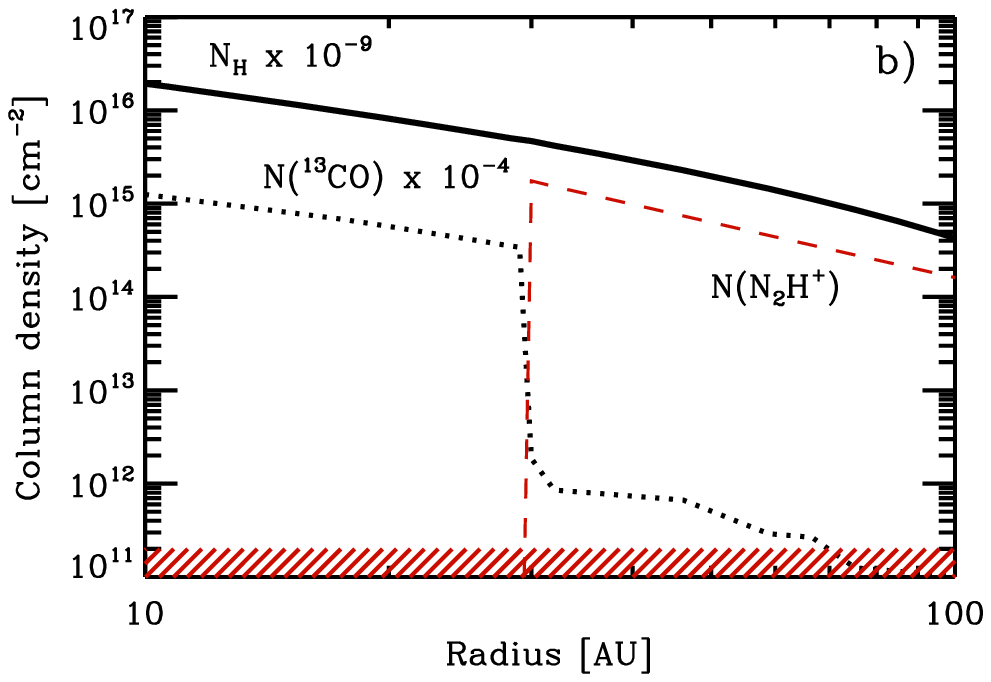}
\end{minipage}\hfill
\includegraphics[width=1.0\textwidth]{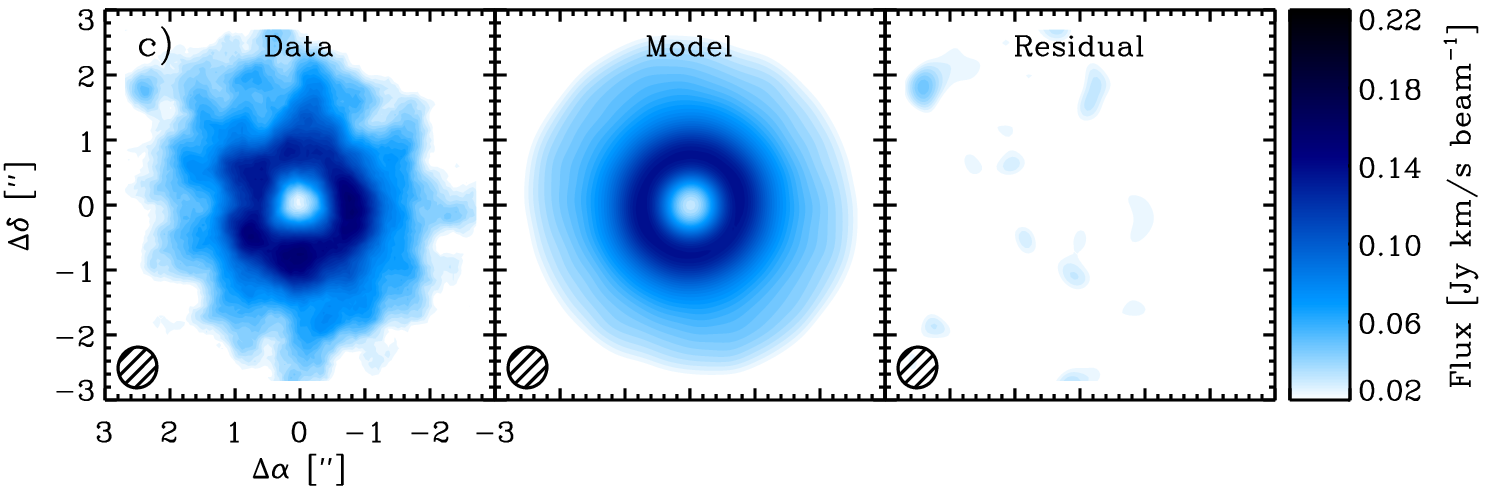}
\caption{Model results for the N$_2$H$^+$ abundance structure toward TW Hya. {\it Upper left:} The $\chi^2$ fit surface  for the power law index and 
inner radius of the N$_2$H$^+$ abundance profile. Contours correspond
to the 1--5 $\sigma$ errors and the blue contour 
marks 3$\sigma$.  {\it Upper right:} The best fit N$_2$H$^+$ column density 
structure, shown together with the total gas column density and the best-fit 
$^{13}$CO column density for CO freeze-out at 17~K. The shaded region
marks the N$_2$H$^+$ 1$\sigma$ detection limit. {\it Lower panel:} 
N$_2$H$^+$ observations, simulated observations of the best-fit N$_2$H$^+$ model, and the imaged residuals, calculated from the visibilities.}
\label{mommap2}
\end{center}
\end{figure} 

\bibliographystyle{Science}

\noindent{\bf Acknowledgments} \\
\noindent We are grateful to S.~Schnee for 
data calibration and reduction assistance. C.Q. would like to thank
the SAO Radio Telescope Data Center (RTDC) staff for their generous
computational support. C.Q., K.I.O. and D.J.W. acknowledges a grant
from NASA Origins of Solar Systems grant
No. NNX11AK63. P.D. acknowledges a grant from PAPIIT-UNAM. 
E.A.B. acknowledges support from NSF Grant\#1008800.
This Report makes use of the following ALMA data:
ADS\/JAO.ALMA\#2011.0.00340.S. ALMA is a partnership of ESO
(representing its member states), NSF (USA), and NINS (Japan),
together with NRC (Canada) and NSC and ASIAA (Taiwan), in cooperation
with the Republic of Chile. The Joint ALMA Observatory is operated by
ESO, AUI/NRAO, and NAOJ. We also make
use of the Submillimeter Array (SMA) data: project \#2004-214 (PI: C.~Qi). 
The SMA is a joint project between the Smithsonian
Astrophysical Observatory and the Academia Sinica Institute of
Astronomy and Astrophysics and is funded by the Smithsonian
Institution and the Academia Sinica.\\

\noindent{\bf Supporting Online Material}\\
\noindent www.sciencemag.org/cgi/content/full/science.1239560/DC1\\
Materials and Methods\\
Table S1--S3\\
Figs S1--S5\\
References (35--52)

\newpage

\section*{Materials and Methods}

\subsection*{Observational details}

Continuum and N$_2$H$^+$ line observations toward TW Hya were carried out
in ALMA band 7 (PI: C. Qi) on 19 November, 2012, with  23 to 26
antennas in the Cycle 0 compact configuration.  
The correlator was configured to observe four windows with a
channel spacing of $\delta\nu$= 61.04 kHz and a bandwidth of 234.375 MHz each. The windows were centered at 372.672 GHz (SPW\#1), the rest frequency of the
N$_2$H$^+$ $J=4-3$ line, 372.421 GHz  (SPW\#0), 358.606 GHz  (SPW\#2), and 357.892 GHz  (SPW\#3) . The nearby quasar J1037-295 was used for phase and gain
calibration and 3C279 and J0522-364 were used as bandpass calibrators. The
primary calibrator Ceres provided a mean flux density of 0.61 Jy for the
gain calibrator J1037-295. The visibility data were reduced and calibrated in CASA 3.4.

The atmospheric transmission in the upper (372 GHz) and
lower (358 GHz) sidebands is very different due to strong
absorption near 370 GHz. Therefore, we reduced the data
separately for both 
sidebands. The continuum visibilities were extracted by
averaging the line-free channels in SPW\# 0, 1 (upper sideband) and 2,3 (lower sideband),
respectively. We carried out self-calibration procedures on the
continuum as demonstrated in the TW Hya Science
Verification Band 7 CASA Guides, 
which are available online
(https://almascience.nrao.edu/alma-data/science-verification/tw-hya).  
The synthesized beam and RMS for the continuum maps are
$0.''63 \times 0.''60$ ($PA=3\degree$), 0.61 mJy
beam$^{-1}$ (upper sideband) and $0.''57 \times 0.''55$ ($PA=17\degree$),
0.24 mJy beam$^{-1}$ (lower sideband). The continuum
peak flux densities
are determined to be 2.0097$\pm$0.0062 Jy at 372 GHz and 1.7814$\pm$
0.0044 Jy at 358 GHz, which agrees with previous SMA observations and ALMA science verification data \cite{Andrews12, Rosenfeld12a}. We applied the upper sideband continuum
self-calibration correction to the N$_2$H$^+$ $J=4-3$ line data and
subtracted the continuum emission in the visibility domain. 

The resulting synthesized beam for the N$_2$H$^+$ $J=4-3$ data cube is
$0.''63 \times 0.''59$ ($PA=-18\degree$), and the 1$\sigma$ rms is
30 mJy beam$^{-1}$ in 0.1 km s$^{-1}$ velocity intervals or 8.1
mJy~beam$^{-1}$~km~s$^{-1}$, which corresponds to a column density
detection limit of 2$\times$10$^{11}$ cm$^{-2}$ at 17 K. 
Fig.~\ref{fig:channelmaps} shows the resulting channel maps for N$_2$H$^+$ $J=4-3$. 

\subsection*{Physical model}

The physical model used to interpret the ALMA observations is a 
steady viscous accretion disk, heated by irradiation from the
central star and mechanical energy generated by viscous dissipation
near the disk midplane\cite{dAlessio98,dAlessio99,dAlessio01,dAlessio06}. The disk model is axisymmetric, in 
vertical hydrostatic equilibrium, and the viscosity follows the
$\alpha$ prescription \cite{Shakura76}. Energy is transported in the disk by 
radiation, convection (in regions where the Schwarzschild stability 
criterion is not satisfied) and a turbulent energy flux. The
penetration of the stellar and shock generated   
radiation is calculated using the two first moments of the radiative
transfer equation, taking into account scattering and absorption by
dust grains. This model framework has been used to successfully
reproduce observed disk structures towards several T Tauri and Herbig
Ae stars \cite{Wilner05,Calvet05,Espaillat07,Qi08,Espaillat10,Qi11}. Following
\cite{Hartmann98,Hughes08, Qi11}, the model includes a tapered
exponential edge to simulate viscous spreading:  $\Sigma \sim \dot{M}
\Omega_k/\alpha T_c$, with $\alpha(R)=\alpha_0 exp(-R/R_c)$. Using the
parameter values listed in Table S1, this physical model provides a
good fit to the broadband spectral energy distribution (SED) of the
disk of TW Hya (Fig.~\ref{fig:sed}), except for the mid- and far-IR wavelengths, which is
complicated by contributions of the inner disk wall, around 3.5--4 AU
and optically thin hot dust from the inner hole, according to
\cite{Calvet02, Uchida04}. The $\alpha_0$ adopted in this model is 
also consistent with the upper limit on the turbulent line widths,
($<$ 40 m s$^{-1}$) at $\sim$1--2
scale heights \cite{Hughes11}, and an ever lower
   turbulence is expected in the midplane.

Following \cite{Qi11}, given a measured disk mass accretion rate,
2$\times$10$^{-9}$ $M_{\odot}$ yr$^{-1}$ \cite{Herczeg04}, and the viscous
$\alpha$ parameter as formulated above, the vertical temperature and
density structures 
are mainly regulated by the degree of grain growth and settling. In
the present model, we introduce this effect in a parametric way. The
dust is assumed to consist of two populations of grains with different
size distribution functions \cite{dAlessio06},  with
$a_{max}^{small}=0.25 \mu$m (as in the interstellar medium) and
$a_{max}^{big}=1$ mm (consistent with the SED slope at mm
wavelengths), and different spatial distributions such that the
abundance of the large grains increases towards the midplane. The
small dust grain to gas mass ratio is parameterized by $\zeta_{\rm small}$,
which is lower than the ISM value because a large fraction of the dust
mass is contained in larger grains. The amount of dust that is in small dust
grains is parameterized by $\epsilon= \zeta_{\rm small}/\zeta_{\rm
  ISM} < 1$. The amount of large dust grains (parameterized by
$\epsilon_{\rm big}$) is calculated so that the total dust mass at
each radius is conserved, and $\epsilon$ is constrained by the slope of
the SED in the  far-IR and sub-mm wavelength range. The maximum grain
size in the model is set to 1 mm, despite observational evidence for
the existence of cm-sized grains \cite{Wilner05},
because we found a power-law size distribution of grains  with a maximum size
in cm could not fit the mm SED slope. This is
probably caused by a radial distribution difference for the mm and cm-sized
dust grains due to differential dust migration. The exclusion of
cm-sized grains should have no effect on the conclusions of this study since
their impact on the midplane temperature structure is minimal. 

The surface that separates the regions
where the small and large dust grain populations dominate is
parameterized by $z_{\rm big}(R)$ in terms of the local gas scale heights, $H$. 
Different $z_{\rm big}$ values are expected to result in disk structures with different thermal profiles because of changes in the shape of the irradiation surface, which determines the
fraction of stellar emission intercepted and reprocessed by the
disk. We vary $z_{\rm big}$ between 2$H$ and 3.5$H$, which results in
the different vertical disk temperature structures seen in
Fig.~\ref{fig:vertical}. Around the observed inner edge of the N$_2$H$^+$ ring,
changing $z_{\rm big}$ also changes the midplane temperature by up to
2 K (Fig.~\ref{fig:midT}). Despite the importance of $z_{\rm big}$ for vertical
temperature and density structure, Fig.~\ref{fig:sed} shows that models with
different $z_{\rm big}$ values present small variations in the SEDs,
and all models fit the observed SED satisfactorily. 
This is generally true for SED modeling because of the degeneracy of
the dust data with the parameter $z_{\rm big}$ \cite{Qi11} and the
nearly face-on disk geometry makes SED modeling especially challenging. 
The details of
the vertical structure in the TW Hya disk are therefore uncertain, and
below we analyze the N$_2$H$^+$ line emission using the full range of
$z_{\rm big}$ values to explore its effect on our conclusions. 

\subsection*{N$_2$H$^+$ line modeling}

We adopted the molecular abundance model introduced by \cite{Qi08,
Qi11}, and  assumed that the N$_2$H$^+$ emission originates in a vertical
layer with a constant abundance between the surface ($\sigma_s$) and
midplane ($\sigma_m$) boundaries which are represented by vertically
integrated hydrogen column densities measured from the disk surface in
units of 1.59$\times$10$^{21}$ cm$^{-2}$. 
We fix the vertical surface boundary $\sigma_s$ to
3.2 and the midplane boundary $\sigma_m$ to 100, which simulates an emission layer close to the midplane, in accordance with model predictions. Fitting these boundaries would require a combination of a well constrained vertical temperature structure and multiple N$_2$H$^+$ transitions and is thus outside of the scope of this study. To test the importance of the assumed vertical structure on the inferred radial distribution of N$_2$H$^+$ we simulate N$_2$H$^+$ visibilities for disk structure models with the range of $z_{\rm big}$ values and disk structures shown in Figs. S2-S4.

We model the radial distribution of N$_2$H$^+$ as a power law
N$_{100}\times(r/100)^p$ with an inner
radius $R_{in}$ and outer radius $R_{out}$, where  $N_{100}$ is the
column density at 100 AU in cm$^{-2}$, $r$ is the distance from the star in
AU, and $p$ is the power-law index. For each $z_{\rm big}$ structure model, we
compute a grid of synthetic N$_2$H$^+$ $J=4-3$ visibility datasets over a range
of $R_{out}$, $R_{in}$, $p$ and $N_{100}$ values and compare with the
observations. The 
best-fit model is obtained by minimizing $\chi^2$, the weighted
difference between the data and the model with the real and imaginary part of the complex
visibility measured in the $(u,v)$-plane sampled by the ALMA
observations. We use the two-dimensional Monte Carlo model RATRAN
\cite{Hogerheijde00}  
to calculate the radiative transfer and molecular excitation. 
The collisional rate coefficients are taken from the Leiden Atomic and
Molecular Database \cite{Schoier05}.

Table~S2 gives the best-fit N$_2$H$^+$ distribution parameters
for each $z_{\rm big}$ model, as well as the corresponding midplane temperature
$T_{c}$ at the N$_2$H$^+$ inner edge. 
The power law index of the surface density varies
  between $2.4$ and $-3.6$, and the column density at the inner edge between
  4$\times$10$^{12}$ and 2$\times$10$^{15}$ cm$^{-2}$. Given the
  1$\sigma$ detection limit of 2$\times$10$^{11}$ cm$^{-2}$, the column density
  contrast at the inner edge of the N$_2$H$^+$ ring is at least 20 and
  could be much larger.
Across this range of models, the
inner radius varies by less than 5~AU, and
the midplane temperature at the inner radius varies by less than
1 K. The inner radius is thus well constrained, which implies that the
key feature of the N$_2$H$^+$ distribution needed to 
constrain the CO snow line is robust with respect to the
details of the physical model assumptions. 
Channel maps for the best-fit model using the fiducial $z_{\rm
  big}=3H$ are shown in Fig.~\ref{fig:channelmaps} together with the observed data and
imaged residuals, demonstrating the excellent agreement.  

\subsection*{CO distribution}

To derive the CO distribution and test the
self-consistency of the fiducial best-fit model, we also modeled two
CO isotopologues which we observed with the Submillimter Array (SMA)
\cite{Ho04} in 2005 February 27 and April 10. The main isotopologue was also observed, but is optically thick and therefore not included in this analysis. The SMA receivers
operated  
in a double-sideband mode with an intermediate frequency (IF) band of 
4--6 GHz from the local oscillator frequency, sent over fiber optic 
transmission lines to 24 overlapping ``chunks'' of the digital
correlator. The correlator was configured to include CO, $^{13}$CO and
C$^{18}$O $J=2-1$ in one setting: the tuning was centered on the
CO $J=2-1$ line at 230.538 GHz in chunk S15, while the $^{13}$CO $J=2-1$
at 220.399 GHz and C$^{18}$O $J=2-1$ at 219.560 GHz were simultaneously
observed in chunk 12 and 22, respectively \cite{Qi06}. Combinations of two array configuration (compact and extended) were used to obtain projected baselines ranging from 6
to 180 m. The observing loops used J1037-295 as the gain
calibrator. The bandpass response was  calibrated using observations
of 3C279. Flux calibration was done using observations of Titan and
Callisto. Routine calibration tasks were performed using the MIR software
package (http://www.cfa.harvard.edu/$\sim$cqi/mircook.html), 
and imaging and deconvolution were accomplished in MIRIAD.  
The integrated fluxes are reported in Table S3. 
Fig.~\ref{fig:cospec} shows the spatially integrated spectra of $^{13}$CO
and C$^{18}$O $J=2-1$ extracted from the SMA channel maps in 8$''$ square
boxes centered on TW~Hya.  

Following \cite{Qi11}, CO is assumed to be present in the disk between a lower
boundary set by the CO freeze-out temperature derived from the N$_2$H$^+$ modeling, and an upper boundary set by photodissociation, though the
choice of upper boundary in this case has a very small effect on the modeled emission profiles
of $^{13}$CO and C$^{18}$O. The CO abundance structure was optimized
using the same procedure as for N$_2$H$^+$ above. Fig.~\ref{fig:cospec} shows that
the best-fit CO abundance distribution fits the CO isotopologue
observations well when using the fiducial disk structure and assuming
standard isotope ratios and CO freeze-out at the N$_2$H$^+$ inner edge
temperature of 17~K. In contrast the models with much smaller $z_{\rm
  big}$ cannot reproduce the relative C$^{18}$O and $^{13}$CO fluxes
without order of magnitude deviations from the cosmic isotope ratios. 
More data with better sensitivity and resolution from the emission
of CO and its isotopologues 
and a detailed surface heating model (as suggested by \cite{Qi06})
are needed to constrain the temperature structure and the $z_{\rm
  big}$ value in the disk of TW Hya.

\clearpage
\makeatletter 
\renewcommand{\thetable}{S\@arabic\c@table}

\begin{table}
\caption{Physical model for the disk of TW Hya \label{tab:model}}
\begin{tabular}{l c }
\hline\hline
Parameters & Values \\
\hline
\multicolumn{2}{c}{Stellar and accretion properties} \\
\hline
Spectral type & K7 \\
Effective temperature: $T_*$ (K) & 4110 \\
Estimated distance: $d$ (pc) & 54 \\
Stellar radius: $R_*$ (R$_\odot$) & 1.04 \\
Stellar mass: $M_*$ (M$_\odot$)& 0.8 \\
Accretion rate: $\dot{M}$ ($M_{\odot}$ yr$^{-1}$) &
2$\times$10$^{-9}$\\
\hline
\multicolumn{2}{c}{Disk structure properties} \\
\hline
Disk mass: $M_d$ (M$_\odot$) & 0.04  \\
Characteristic radius: $R_c$ (AU) & 60 \\
Viscosity coefficient: $\alpha_0$ & 0.0007 \\
Depletion factor of the atmospheric small grains: $\epsilon^a$ & 0.01 \\
$z_{\rm big}$$^a$ (H$^b$) & 3.0 \\
\hline
\multicolumn{2}{c}{Disk geometric and kinematic properties$^c$} \\
\hline
Inclination: $i$ (deg) & 6 \\
Systemic velocity: $V_{\rm LSR}$ (km s$^{-1}$) & 2.86 \\
Turbulent line width: $\delta$v$_{turb}$ (km s$^{-1}$) &0.05 \\
Position angle: $P.A.$ (deg) & 155 \\
\hline
$^{a}$See definition in paper.\\
$^{b}$Gas scale height.\\
$^{c}$Parameters adopted from \cite{Qi04, Qi06, Qi08, Andrews12}.\\
\end{tabular}
\end{table}

\begin{table}
\caption{N$_2$H$^+$ $J=4-3$ fitting results$^a$}
\begin{center}
\begin{tabular}{l c c c c c}
\hline
\hline
$z_{\rm big}$ ($H$)& $R_{in}$ (AU) & $T_{\rm c}$$^b$ (K)& $p$ &
$N_{100}$ (cm$^{-2}$) & $R_{out}$ (AU) \\
\hline
2.0 & 25$^{+4}_{-6}$ & 15--18 &  2.4$^{+0.6}_{-0.3}$   & (1.4$\pm$0.2) $\times$10$^{14}$ & 150$\pm$10 \\
2.5 & 30$^{+1}_{-3}$ & 15--16 &  0.4$^{+0.6}_{-0.4}$  & (2.9$\pm$0.5) $\times$10$^{14}$ & 140$\pm$10 \\
3.0 & 30$^{+1}_{-2}$ & 16--17 & $-$2.0$^{+0.5}_{-0.7}$  & (1.6$\pm$0.3) $\times$10$^{14}$ & 140$\pm$10 \\
3.5 & 30$^{+1}_{-4}$ & 16--18 & $-$3.6$^{+0.6}_{-0.8}$  & (2.5$\pm$0.4) $\times$10$^{13}$ & 140$\pm$10 \\
\hline
$^{a}$Errors within 3$\sigma$.\\
$^{b}$Temperature range based on Fig.~\ref{fig:midT}.\\
\end{tabular}
\end{center}
\label{tab:fitting}
\end{table}   

\begin{table}
\caption{TW Hya CO isotopologue observation results.}
\begin{center}
\begin{tabular}{l c c c}
\hline
\hline
Transition&Frequency (GHz)& Beam /$P.A.$ & $\int F dv$ (Jy km s$^{-1}$)
\\
\hline
$^{13}$CO $J=2-1$&220.399& $2.''7 \times 1.''8$ / $-$3.0\degree & 2.72[0.18]\\
C$^{18}$O $J=2-1$&219.560& $2.''8 \times 1.''9$ / $-$1.3\degree & 0.68[0.18]\\
\hline
\end{tabular}
\end{center}
\label{tab:line}
\end{table}

\clearpage
\setcounter{figure}{0}
\makeatletter 
\renewcommand{\thefigure}{S\@arabic\c@figure} 

\begin{figure}[htbp]
\centering
\includegraphics[width=6in]{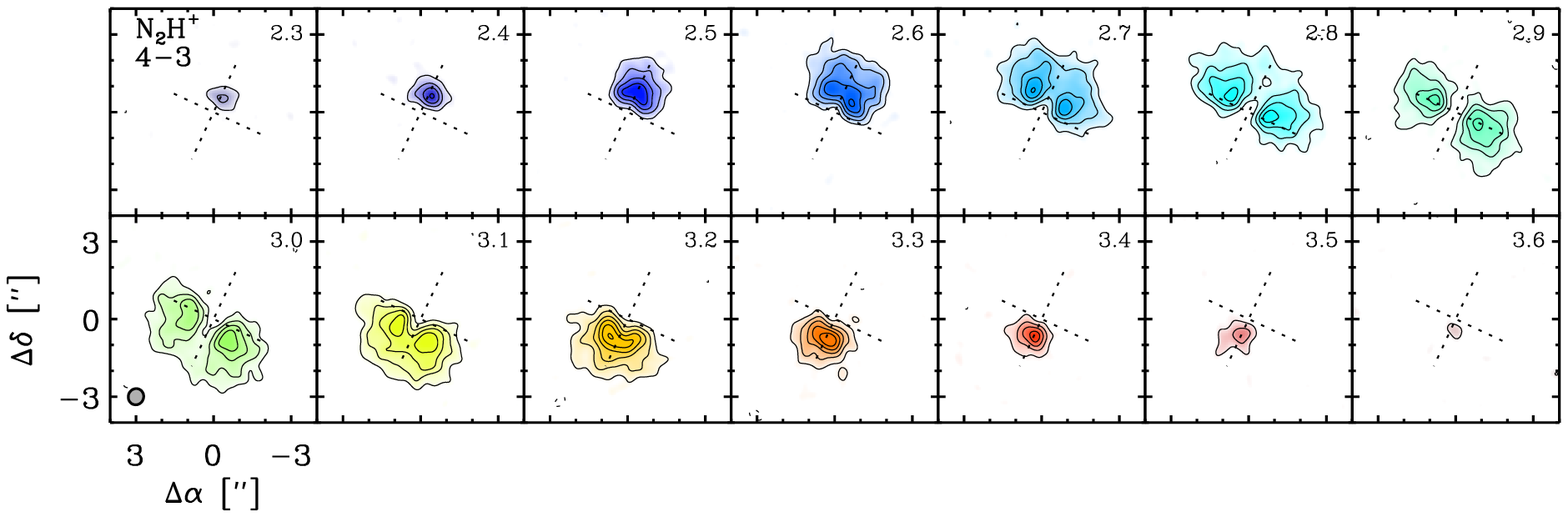}
\vskip 5mm
\includegraphics[width=6in]{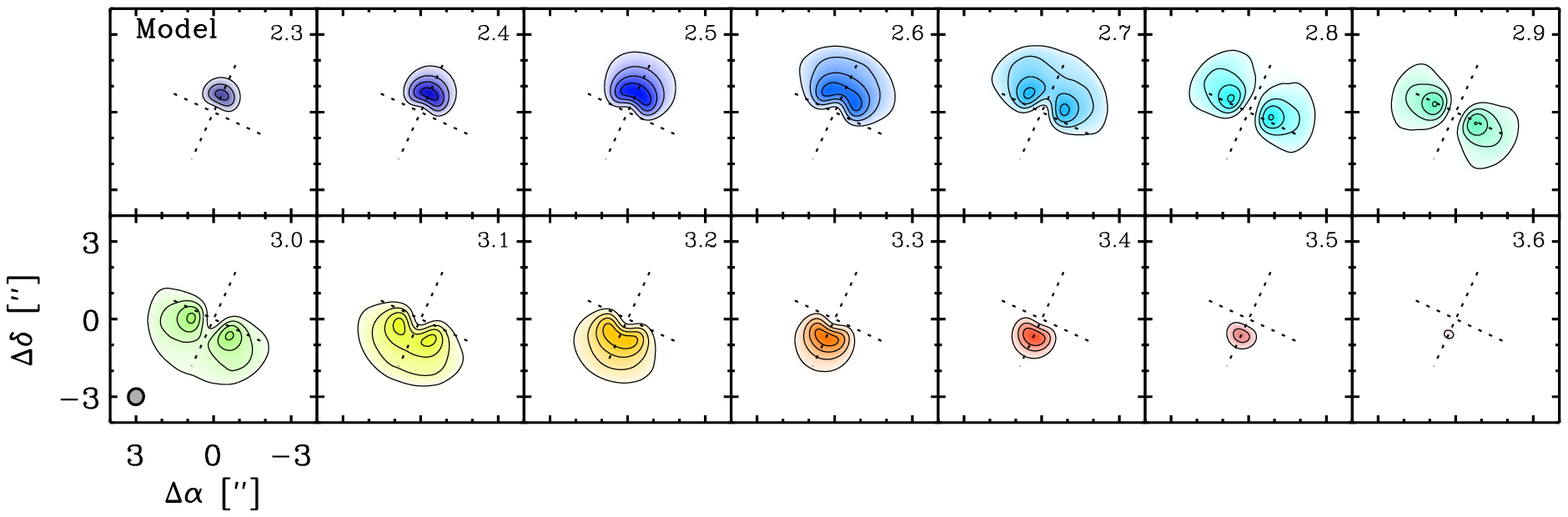}
\vskip 5mm
\includegraphics[width=6in]{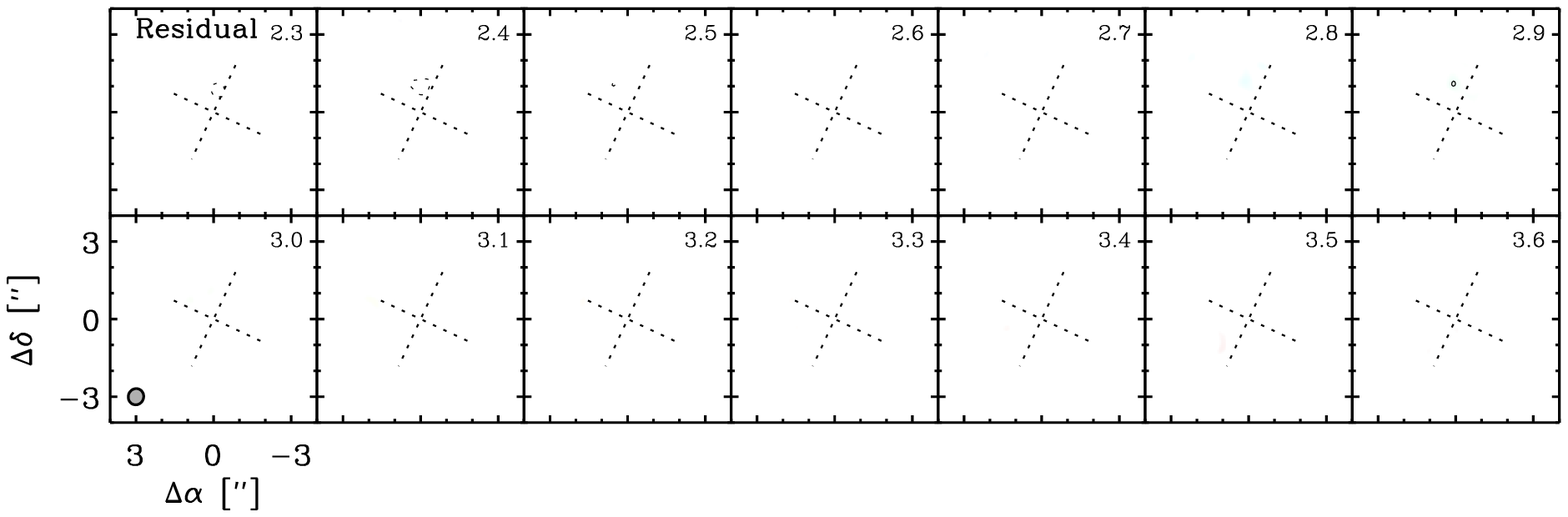}
\caption{ Channel maps of the N$_2$H$^+$ $J=4-3$ line emission observed by ALMA from the disk
  around TW~Hya. The LSR velocity is indicated in the upper right
  of each channel, while the synthesized beam size and orientation
  ($0.''63 \times 0.''59$ at a position angle of $-$18.1$^{\circ}$)
  is indicated in the lower left  
  panel. The contours are 0.03 (1$\sigma$) $\times
  [3,6,9,12,15,18]$ Jy beam$^{-1}$ .}
 \label{fig:channelmaps}
\end{figure}

\begin{figure}[htbp]
\centering
\includegraphics[width=6in]{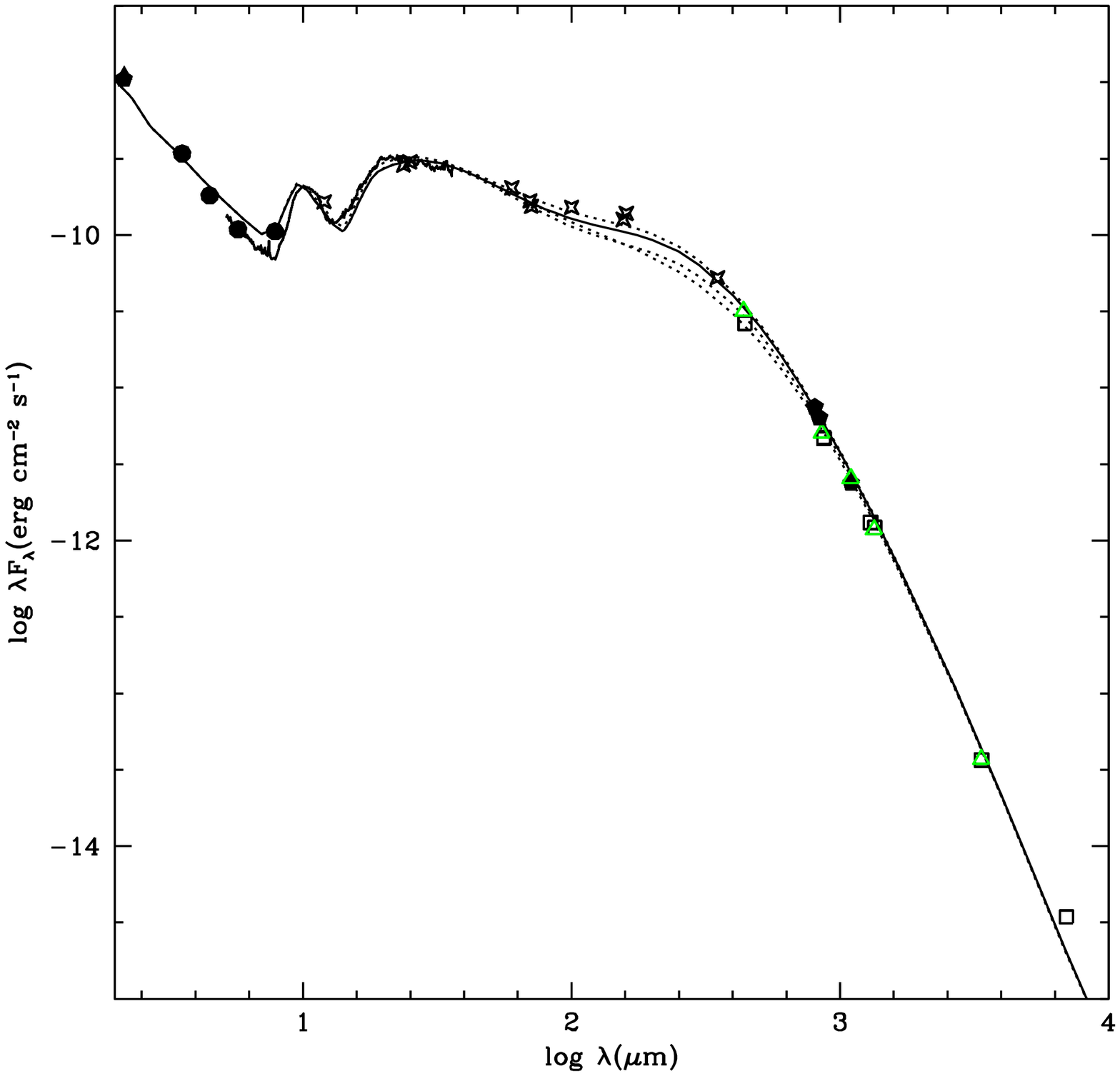}
\caption{The TW~Hya SED and the model results for
$\dot{M}=2  \times 10^{-9} \MSUNYR$, $\alpha_0=0.0007$,
  $\epsilon=0.01$, and $R_c=60$ AU. See the SED references in
    \cite{Andrews12}. The new mm/submm fluxes are from the ALMA
    science verification data and this paper (marked by green
    triangles). 
The different model SED lines correspond to $z_{\rm big}/H= 2-3.5$, with the fiducial 3H model shown with a solid line, demonstrating that the SED modeling does not provide strong constraints on this parameter.  }
\label{fig:sed}
\end{figure}

\begin{figure}[htbp]
\centering
\includegraphics[width=5in]{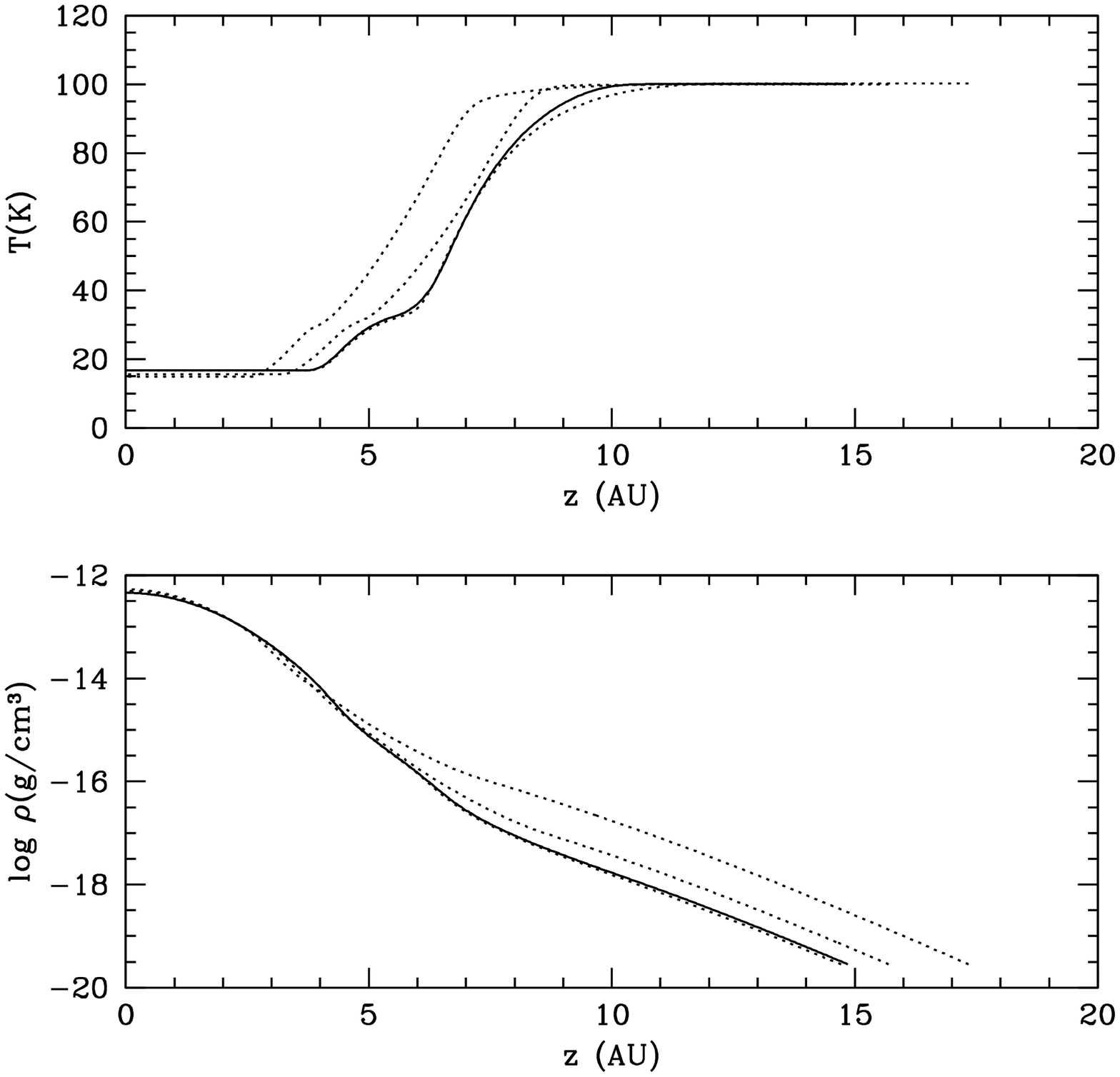}
\caption{Vertical temperature and density profiles at $R=30$ AU for
  disk models with $\dot{M}=2 \times 10^{-9} \ \MSUNYR$, $\alpha_0=0.0007$, 
$\epsilon=0.01$, $R_c=60$ AU, and different 
values of $z_{\rm big}$ values. {\sl Upper panel:} Temperature versus height at $R=30$ AU for 
disk models with different 
values of $z_{\rm big}/H=$  2.0, 2.5, 3.0, 3.5 (from left to
right). The fiducial model, with $z_{\rm big}=3 H$ is shown with a solid line.
{\sl Lower panel:} Density versus height at $R=30$ AU for the same models. The
lines from top to bottom correspond to models with $z_{\rm big}/H=$  2.0, 2.5, 3.0, 3.5.
}
\label{fig:vertical}
\end{figure}

\begin{figure}[htp]
\centering
\includegraphics[width=4in]{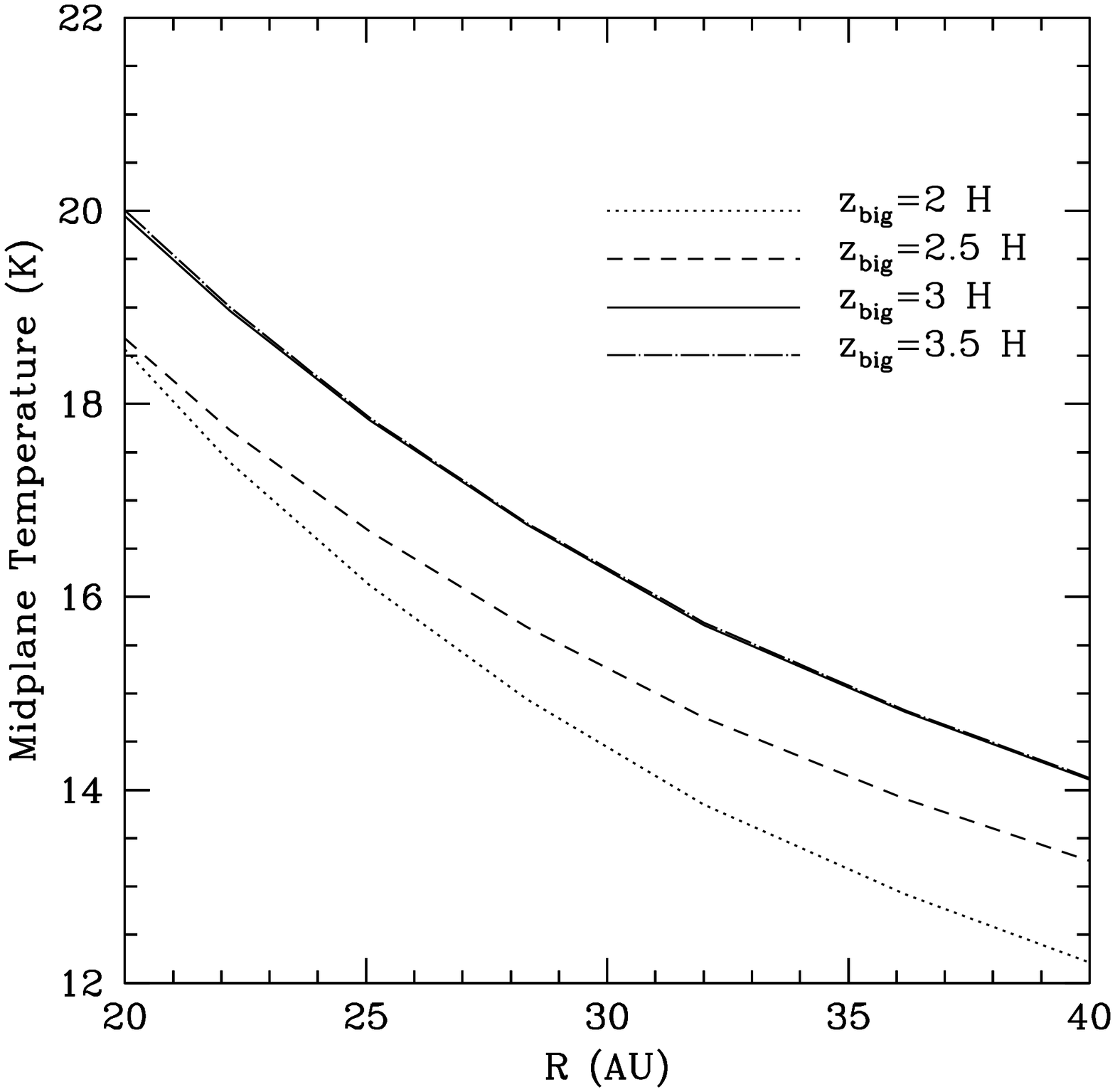}
\caption{Midplane temperature profiles for the TW Hya disk for different $z_{\rm big}$ values showing the effect of $z_{\rm big}$ on the midplane temperature around the N$_2$H$^+$ inner edge. \label{fig:midT}}
\end{figure}

\begin{figure}[htp]
\centering
\includegraphics[width=3in]{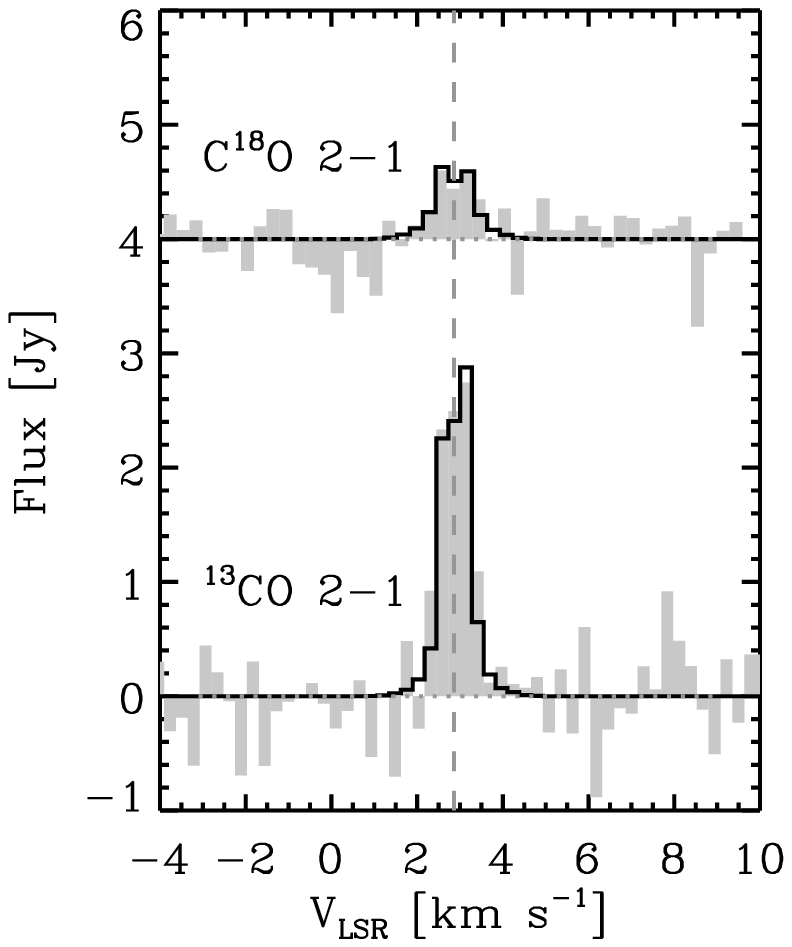}
\caption{CO isotopologue lines toward TW Hya, observed with the SMA
  (grey) and the best-fit CO modeling results using the fiducial disk
  structure developed to interpret the N$_2$H$^+$ observations. The
  dashed line marks the $V_{\rm LSR}$ toward TW~Hya. \label{fig:cospec}}
\end{figure}


\begin{thebibliography}{52}

\bibitem{Lewis74}
J.~S. {Lewis}, {\it Science\/} {\bf 186}, 440 (1974).

\bibitem{Stevenson88}
D.~J. {Stevenson}, J.~I. {Lunine}, {\it Icarus\/} {\bf 75}, 146 (1988).

\bibitem{Ciesla06}
F.~J. {Ciesla}, J.~N. {Cuzzi}, {\it Icarus\/} {\bf 181}, 178 (2006).

\bibitem{Johansen07}
A.~{Johansen}, {\it et~al.\/}, {\it Nature\/} {\bf 448}, 1022 (2007).

\bibitem{Chiang10}
E.~{Chiang}, A.~N. {Youdin}, {\it Annual Review of Earth and Planetary
  Sciences\/} {\bf 38}, 493 (2010).

\bibitem{Gundlach11}
B.~{Gundlach}, S.~{Kilias}, E.~{Beitz}, J.~{Blum}, {\it Icarus\/} {\bf 214},
  717 (2011).

\bibitem{Ros13}
K.~{Ros}, A.~{Johansen}, {\it A\&A\/} {\bf 552}, A137 (2013).

\bibitem{Oberg11e}
K.~I. {{\"O}berg}, R.~{Murray-Clay}, E.~A. {Bergin}, {\it ApJL\/} {\bf 743},
  L16 (2011).

\bibitem{Hayashi81}
C.~{Hayashi}, {\it Progress of Theoretical Physics Supplement\/} {\bf 70}, 35
  (1981).

\bibitem{Herbst09}
E.~{Herbst}, E.~F. {van Dishoeck}, {\it ARA\&A\/} {\bf 47}, 427 (2009).

\bibitem{Wilner05}
D.~J. {Wilner}, P.~{D'Alessio}, N.~{Calvet}, M.~J. {Claussen}, L.~{Hartmann},
  {\it ApJL\/} {\bf 626}, L109 (2005).

\bibitem{Bisschop06}
S.~E. {Bisschop}, H.~J. {Fraser}, K.~I. {{\"O}berg}, E.~F. {van Dishoeck},
  S.~{Schlemmer}, {\it A\&A\/} {\bf 449}, 1297 (2006).

\bibitem{Aikawa99}
Y.~{Aikawa}, E.~{Herbst}, {\it A\&A\/} {\bf 351}, 233 (1999).

\bibitem{Qi11}
C.~{Qi}, {\it et~al.\/}, {\it ApJ\/} {\bf 740}, 84 (2011).

\bibitem{Qi13}
C.~{Qi}, K.~I. {{\"O}berg}, D.~J. {Wilner}, {\it ApJ\/} {\bf 765}, 34 (2013).

\bibitem{Bergin02}
E.~A. {Bergin}, J.~{Alves}, T.~{Huard}, C.~J. {Lada}, {\it ApJL\/} {\bf 570},
  L101 (2002).

\bibitem{Jorgensen04}
J.~K. {J{\o}rgensen}, {\it A\&A\/} {\bf 424}, 589 (2004).

\bibitem{Oberg11a}
K.~I. {{\"O}berg}, {\it et~al.\/}, {\it ApJ\/} {\bf 734}, 98 (2011).

\bibitem{Walsh12}
C.~{Walsh}, H.~{Nomura}, T.~J. {Millar}, Y.~{Aikawa}, {\it ApJ\/} {\bf 747},
  114 (2012).

\bibitem{SOM}
Materials and methods are available as supporting material on Science Online.

\bibitem{Kastner97}
J.~H. {Kastner}, B.~{Zuckerman}, D.~A. {Weintraub}, T.~{Forveille}, {\it
  Science\/} {\bf 277}, 67 (1997).

\bibitem{Qi08}
C.~{Qi}, D.~J. {Wilner}, Y.~{Aikawa}, G.~A. {Blake}, M.~R. {Hogerheijde}, {\it
  ApJ\/} {\bf 681}, 1396 (2008).

\bibitem{Bergin13}
E.~A. {Bergin}, {\it et~al.\/}, {\it Nature\/} {\bf 493}, 644 (2013).

\bibitem{Andrews12}
S.~M. {Andrews}, {\it et~al.\/}, {\it ApJ\/} {\bf 744}, 162 (2012).

\bibitem{Rosenfeld12a}
K.~A. {Rosenfeld}, {\it et~al.\/}, {\it ApJ\/} {\bf 757}, 129 (2012).

\bibitem{Hollenbach09}
D.~{Hollenbach}, M.~J. {Kaufman}, E.~A. {Bergin}, G.~J. {Melnick}, {\it ApJ\/}
  {\bf 690}, 1497 (2009).

\bibitem{Willacy07}
K.~{Willacy}, {\it ApJ\/} {\bf 660}, 441 (2007).

\bibitem{Hogerheijde00}
M.~R. {Hogerheijde}, F.~F.~S. {van der Tak}, {\it A\&A\/} {\bf 362}, 697
  (2000).

\bibitem{Lecar06}
M.~{Lecar}, M.~{Podolak}, D.~{Sasselov}, E.~{Chiang}, {\it ApJ\/} {\bf 640},
  1115 (2006).

\bibitem{DodsonRobinson09}
S.~E. {Dodson-Robinson}, K.~{Willacy}, P.~{Bodenheimer}, N.~J. {Turner}, C.~A.
  {Beichman}, {\it Icarus\/} {\bf 200}, 672 (2009).

\bibitem{Tsiganis05}
K.~{Tsiganis}, R.~{Gomes}, A.~{Morbidelli}, H.~F. {Levison}, {\it Nature\/}
  {\bf 435}, 459 (2005).

\bibitem{Owen93}
T.~C. {Owen}, {\it et~al.\/}, {\it Science\/} {\bf 261}, 745 (1993).

\bibitem{Tegler12}
S.~C. {Tegler}, {\it et~al.\/}, {\it ApJ\/} {\bf 751}, 76 (2012).

\bibitem{Mumma11}
M.~J. {Mumma}, S.~B. {Charnley}, {\it ARA\&A\/} {\bf 49}, 471 (2011).

\bibitem{dAlessio98}
P.~{D'Alessio}, J.~{Canto}, N.~{Calvet}, S.~{Lizano}, {\it ApJ\/} {\bf 500},
  411 (1998).

\bibitem{dAlessio99}
P.~{D'Alessio}, N.~{Calvet}, L.~{Hartmann}, S.~{Lizano}, J.~{Cant{\'o}}, {\it
  ApJ\/} {\bf 527}, 893 (1999).

\bibitem{dAlessio01}
P.~{D'Alessio}, N.~{Calvet}, L.~{Hartmann}, {\it ApJ\/} {\bf 553}, 321 (2001).

\bibitem{dAlessio06}
P.~{D'Alessio}, N.~{Calvet}, L.~{Hartmann}, R.~{Franco-Hern{\'a}ndez},
  H.~{Serv{\'{\i}}n}, {\it ApJ\/} {\bf 638}, 314 (2006).

\bibitem{Shakura76}
N.~I. {Shakura}, R.~A. {Sunyaev}, {\it MNRAS\/} {\bf 175}, 613 (1976).

\bibitem{Calvet05}
N.~{Calvet}, {\it et~al.\/}, {\it ApJL\/} {\bf 630}, L185 (2005).

\bibitem{Espaillat07}
C.~{Espaillat}, {\it et~al.\/}, {\it ApJL\/} {\bf 670}, L135 (2007).

\bibitem{Espaillat10}
C.~{Espaillat}, {\it et~al.\/}, {\it ApJ\/} {\bf 717}, 441 (2010).

\bibitem{Hartmann98}
L.~{Hartmann}, N.~{Calvet}, E.~{Gullbring}, P.~{D'Alessio}, {\it ApJ\/} {\bf
  495}, 385 (1998).

\bibitem{Hughes08}
A.~M. {Hughes}, D.~J. {Wilner}, C.~{Qi}, M.~R. {Hogerheijde}, {\it ApJ\/} {\bf
  678}, 1119 (2008).

\bibitem{Calvet02}
N.~{Calvet}, {\it et~al.\/}, {\it ApJ\/} {\bf 568}, 1008 (2002).

\bibitem{Uchida04}
K.~I. {Uchida}, {\it et~al.\/}, {\it ApJS\/} {\bf 154}, 439 (2004).

\bibitem{Hughes11}
A.~M. {Hughes}, D.~J. {Wilner}, S.~M. {Andrews}, C.~{Qi}, M.~R. {Hogerheijde},
  {\it ApJ\/} {\bf 727}, 85 (2011).

\bibitem{Herczeg04}
G.~J. {Herczeg}, B.~E. {Wood}, J.~L. {Linsky}, J.~A. {Valenti}, C.~M.
  {Johns-Krull}, {\it ApJ\/} {\bf 607}, 369 (2004).

\bibitem{Schoier05}
F.~L. {Sch{\"o}ier}, F.~F.~S. {van der Tak}, E.~F. {van Dishoeck}, J.~H.
  {Black}, {\it A\&A\/} {\bf 432}, 369 (2005).

\bibitem{Ho04}
P.~T.~P. {Ho}, J.~M. {Moran}, K.~Y. {Lo}, {\it ApJL\/} {\bf 616}, L1 (2004).

\bibitem{Qi06}
C.~{Qi}, {\it et~al.\/}, {\it ApJL\/} {\bf 636}, L157 (2006).

\bibitem{Qi04}
C.~{Qi}, {\it et~al.\/}, {\it ApJL\/} {\bf 616}, L11 (2004).

\end{thebibliography}
\end{document}